\documentclass[prd, preprint,preprintnumbers,floatfix, superscriptaddress,nofootinbib] {revtex4-1}
\usepackage{epsfig}
\usepackage{subfigure}
\usepackage{dcolumn}
\usepackage{bm}
\usepackage[usenames ,dvipsnames]{xcolor}
\usepackage{slashed}
\usepackage{graphicx,color}
\usepackage{hyperref}
\usepackage{amsmath}
\usepackage{amssymb}
\usepackage[normalem]{ulem}

\newcommand{\vsig}{\mbox{\boldmath$\sigma$\unboldmath}}

\begin{document}

\title{Investigating $\Omega_c$ spectroscopy in two-body $\Omega_b$ decays}

\author{Juan Wang}\email{wjuanmm@163.com}
\affiliation{School of Physics and Information Engineering,
Shanxi Normal University, Taiyuan 030031, China}

\author{Kai-Lei Wang}\email{wangkaileicz@foxmail.com}
\affiliation{Department of Physics,
Changzhi University, Changzhi, Shanxi 046011, China}

\author{Yu-Kuo Hsiao}\email{yukuohsiao@gmail.com}
\affiliation{School of Physics and Information Engineering,
Shanxi Normal University, Taiyuan 030031, China}

\date{\today}

\begin{abstract}
We investigate the $1S$-, $1P$-, $2S$-, and $1D$-wave $\Omega_c(css)$ spectroscopy
through the non-leptonic decays of $\Omega_b^-$ baryon within the constituent quark model.
For the lowest-lying $1S$-wave $\Omega_c$ state, we obtain the branching fractions
${\cal B}(\Omega_b^- \to \Omega_c^0 \pi^-,\Omega_c^0 \rho^-)=(1.2,6.3) \times 10^{-3}$,
which are consistent with existing model predictions.
For $\Omega_c(3000)$, $\Omega_c(3050)$, $\Omega_c(3065)$,
and $\Omega_c(3090)$, observed in the proton-proton and $e^+ e^-$ collisions and
interpreted as members of the $1P$-wave multiplet (collectively denoted as $\Omega_c^{**}$),
we predict the branching fractions of $\Omega_b^-\to\Omega_c^{**}\pi^-,\Omega_c^{**}\rho^-$
at the level of $10^{-3}$. Assigning the newly observed $\Omega_c(3327)$ baryon
to a $1D$-wave excitation with $J^P=5/2^+$ or $7/2^+$, we obtain
${\cal B}[\Omega_b^-\to \Omega_c(3327)^0\pi^-,\Omega_c(3327)^0\rho^-]
=(2.0,5.7)\times 10^{-3}$ or $(4.6,0.8)\times 10^{-3}$, respectively.
The pronounced differences between these two scenarios
provide a clear discriminant that can be tested in future measurements at LHCb.
\end{abstract}


\maketitle

\section{Introduction}
A singly charmed $\Omega_c$ baryon is composed of three quarks, $css$,
in which the two strange quarks form a correlated diquark subsystem.
The orbital motion between the charm quark and the light diquark plays a central role
in shaping the $\Omega_c$ spectrum, while possible internal orbital excitations of the diquark
may also contribute. In addition, the spin of the $ss$ diquark and the spin of the charm quark
provide further degrees of freedom, whose couplings with the orbital angular momentum
give rise to the observed pattern of spin-parity states.
Within the framework of heavy-quark effective theory,
a systematic classification of the $\Omega_c$ spectrum has been established
~\cite{Chen:2022asf,Roberts:2007ni,Yoshida:2015tia,Luo:2025sns}.
Experimental measurements therefore play a key role in validating and refining
the established $\Omega_c$ spectroscopy.

Several $\Omega_c$ baryons have been observed in the proton-proton collision by LHCb~\cite{LHCb:2017uwr,LHCb:2023sxp}
and the $e^+ e^-$ collision by Belle~\cite{Belle:2017ext}.
However, not all of the observed states have been unambiguously assigned definite quantum numbers~\cite{LHCb:2021ptx}.
For instance, the $\Omega_c(3327)$ baryon, newly observed in Ref.~\cite{LHCb:2023sxp},
has been interpreted as a $1D$ state in several theoretical studies~\cite{Zhong:2025oti,Wang:2023wii,Luo:2023sra,
Pan:2023hwt,Yu:2023bxn,Garcia-Tecocoatzi:2024aqz}, while its spin-parity
quantum numbers $J^P$ remain undetermined.
The excited $\Omega_c$ states observed in Refs.~\cite{LHCb:2017uwr,Belle:2017ext}
comprise five states in total and collectively denoted as $\Omega_c^{**}$ by LHCb~\cite{LHCb:2017uwr}.
These states have been extensively investigated within a wide variety of theoretical frameworks
~\cite{Kucukyilmaz:2025rsd,Li:2025frt,Weng:2024roa,Li:2024zze,
Karliner:2017kfm,Garcia-Tecocoatzi:2022zrf,Karliner:2023okv,Ortiz-Pacheco:2023kjn,
Bahtiyar:2020uuj,Luo:2023sra,Garcia-Tecocoatzi:2024aqz,
Pan:2023hwt,Yu:2023bxn,
Yu:2022ymb,Chen:2017sci,
Agaev:2017lip,Oudichhya:2023awb,Jakhad:2023mni,
Padmanath:2017lng,Agaev:2017jyt,Wang:2017zjw,
Aliev:2017led,Agaev:2017lip,Wang:2017xam,Chen:2017sci,
Aliev:2018uby,Cheng:2017ove,Oudichhya:2021kop,Oudichhya:2023awb,
Wang:2017hej,Wang:2017kfr,Yao:2018jmc,Wang:2017vnc,
Ali:2017wsf,Chen:2017gnu,Zhao:2017fov,Garcia-Tecocoatzi:2022zrf,Kim:2017khv,
Santopinto:2018ljf,Jia:2020vek,Luo:2023sra}.
Among the $\Omega_c^{**}$ states,
$\Omega_c(3000)$, $\Omega_c(3050,3065)$, $\Omega_c(3090)$ are commonly interpreted
as members of the $1P$ multiplet with quantum numbers $J^P=1/2^-$, $3/2^-$, and $5/2^-$, respectively,
while the quantum numbers of $\Omega_c(3120)$ remain uncertain.
Alternatively, the $\Omega_c^{**}$ states have also been proposed as more exotic configurations,
such as hadronic molecules or pentaquark states~\cite{
Ozdem:2024ydl,
Wang:2018alb,Wang:2021cku,Yang:2017rpg,Huang:2017dwn,An:2017lwg,
Anisovich:2017aqa,Nieves:2017jjx,Liu:2018bkx,Chen:2017xat,Montana:2018teh,
Montana:2017kjw,Ramos:2020bgs,Praszalowicz:2022hcp,
Kim:2017jpx,Debastiani:2017ewu,Debastiani:2018adr,Wang:2017smo,Huang:2018wgr,Zhu:2022fyb,Xin:2023gkf,Yan:2023ttx,Yan:2023tvl,Ozdem:2023okg,Feng:2023ixl}.
Clearly, further theoretical and experimental investigations are required to clarify
the nature and quantum-number assignments of these excited $\Omega_c$ baryons.

The LHCb Collaboration has recently observed the resonant decays
$\Omega_b^-\to \Omega_c^{**}\pi^-, \Omega_c^{**}\to \Xi_c^+ K^-$~\cite{LHCb:2021ptx},
in which the states $\Omega_c(3000)$, $\Omega_c(3050)$, $\Omega_c(3065)$, and $\Omega_c(3090)$
reappear as resonance structures in the $\Xi_c^+ K^-$ invariant mass spectrum. As a result,
the existence of these states, established in the $pp$ and $e^+ e^-$ collisions,
is independently confirmed in weak decays. This observation opens up a new avenue
for investigating the $\Omega_c$ spectroscopy, while simultaneously raising several important questions.
First, the $\Omega_c(3120)$ baryon remains the only $\Omega_c^{**}$ state
that has not been observed in this decay mode.
Second, the $\Omega_c(3327)$ state, as well as other higher excited $\Omega_c$ baryons,
can in principle also be produced in $\Omega_b^-$ decays; however,
no corresponding signals were reported until very recently. Third, although four $\Omega_c^{**}$
states have been re-established in weak decays, their branching fractions have not yet been measured.
In view of these issues, we consider the decays $\Omega_b^-\to \Omega_c^{(*)0}M$
as a suitable test ground. These channels involve the lowest-lying $1S$ baryon $\Omega_c^0$,
as well as the excited $1P$- and $1D$-wave states.
The final-state meson $M$, where $M$ can be $\pi$ or $\rho$, provides
scalar or vector quantum numbers that couple differently to $\Omega_c^{(*)0}$ states.

Theoretical studies of semileptonic and charmful $\Omega_b$ decays~\cite{Ortiz-Pacheco:2020hmj,Hsiao:2021mlp,Wang:2024mjw,Duan:2025aur},
as well as the two-body processes $\Omega_b\to \Omega_c M$~\cite{Cheng:1996cs,Ivanov:1997hi,
Ivanov:1997ra,Gutsche:2018utw,Zhao:2018zcb,Neishabouri:2026mjn,Patel:2025gbw},
remain rather limited in the literature. In particular, the decays
$\Omega_b\to \Omega_c^{*}M$ have been much less extensively
explored~\cite{Li:2021kfb,Chua:2018lfa,Chua:2019yqh}.
Motivated by this situation,
we investigate the decays $\Omega_b^-\to \Omega_c^{(*)0}M$
within the non-relativistic constituent quark model (CQM).
The CQM has proven to be a reliable framework in describing hadronic decays of hyperons~\cite{Niu:2020aoz},
$D$ mesons~\cite{Cao:2023csx,Cao:2023gfv}, and heavy baryons containing $b$ or $c$ quarks~\cite{Niu:2020gjw,Wang:2022zja,Niu:2021qcc,Pervin:2006ie,Pervin:2005ve}.
These processes therefore provide useful probes for exploring the $1S$-, $1P$-, $2S$-,
and $1D$-wave structures in the $\Omega_c$ spectrum.

\section{Formalism}
The decays $\Omega_b^-\to \Omega_c^{(*)0}M$ with $M=(\pi^-,\rho^-)$
proceed through the quark-level $b\to c\bar{u}d$ transition.
The according effective Hamiltonian is given by~\cite{Buchalla:1995vs}
\begin{eqnarray}\label{dww1}
{\cal H}_{eff}=\frac{G_F}{\sqrt{2}}V_{cb} V_{ud}^{\ast} (c_1 O_1+ c_2 O_2)\,,
\end{eqnarray}
where $G_F$ is the Fermi constant, $c_{1,2}$ are the Wilson coefficients,
and $V_{cb}(V_{ud}^*)$ the Cabibbo-Kobayashi-Maskawa (CKM) matrix elements.
Explicitly, the current-current operators $O_{1,2}$ read
\begin{eqnarray}\label{dww2}
O_1&=&
\bar{\psi}_{\bar{u}_\beta}\gamma_\mu(1-\gamma_5)\psi_{d_\beta}
\bar{\psi}_{\bar{c}_\alpha}\gamma^\mu(1-\gamma_5)\psi_{b_\alpha}\,,\nonumber\\
O_2&=&
\bar{\psi}_{\bar{u}_\alpha}\gamma_\mu(1-\gamma_5)\psi_{d_\beta}
\bar{\psi}_{\bar{c}_\beta}\gamma^\mu(1-\gamma_5)\psi_{b_\alpha},
\end{eqnarray}
where $\psi_{j_\delta}$ represent the $j$-quark field $[j=(u,d,c,b)]$
and $\delta=(\alpha,\beta)$ denotes a color quantum number.
The amplitude of $\Omega_b^-\to \Omega_c^{(*)0}M$
through ${\cal H}_{eff}$ in Eq.~(\ref{dww1})
is derived as~\cite{Hsiao:2015txa,Hsiao:2015cda,Hsiao:2021mlp}
\begin{eqnarray}\label{dww3}
{\cal M}
=\frac{G_F}{\sqrt{2}}V_{cb} V_{ud}^{\ast} a_1
\langle M\Omega_c^{(*)0}|\bar{\psi}_{\bar{u}_\beta}\gamma_\mu(1-\gamma_5)\psi_{d_\beta}
\bar{\psi}_{\bar{c}_\alpha}\gamma^\mu(1-\gamma_5)\psi_{b_\alpha}|\Omega_b^-\rangle\,,
\end{eqnarray}
with $a_1=c_1^{eff}+c_2^{eff}/N_c^{eff}$, where $c_{1,2}^{eff}$ denote the effective Wilson coefficient
that receive the quark-rescattering effects, and $N_c^{eff}$ denote the effective color number.
By shifting $N_c^{eff}$ from $N_c=3$, the non-factorizable QCD loop effects
can be estimated~\cite{ali,Hsiao:2014mua}.
In Eq.~(\ref{dww3}), the two currents in $O_1$ are split into the two matrix elements,
and $O_2\simeq O_1/N_c^{(eff)}$ as a result of the factorization approximation has been combined as
part of the $O_1$ contribution.

\begin{figure}[t]
\centering
\includegraphics[width=0.45\textwidth]{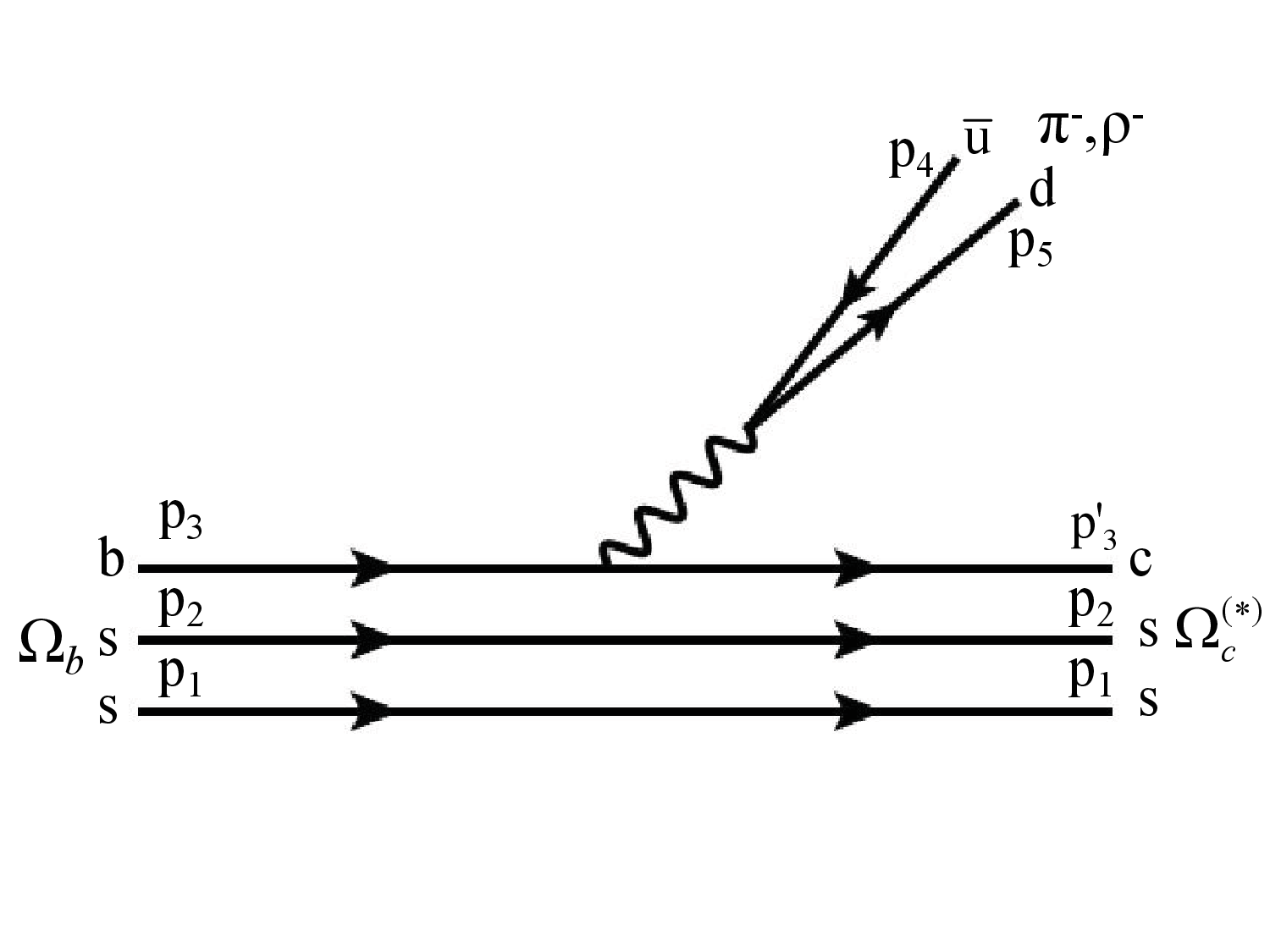}
\vspace{-1.3 cm}
\caption{Feynman diagram for the processes
of $\Omega_b \rightarrow \Omega_c^{(\ast)} {M}$.}\label{tu}
\end{figure}
The amplitude in Eq.~(\ref{dww3}) can be further derived,
where $O_1$ being split into the two matrix elements is presented as
$O_1\simeq O_{\rm w}^{PC}+O_{\rm w}^{PV}$ in the non-relativistic framework.
The parity-conserving and parity-violating operators of the weak transition,
denoted by $O_{\rm w}^{PV}$ and $O_{\rm w}^{PV}$, respectively,
are given by
\begin{eqnarray}\label{Hw}
O_{\rm w}^{PC}&=&
\delta^3(\textbf{p}_3-\textbf{p}_3'-\textbf{p}_4-\textbf{p}_5)/(2\pi)^3\hat{O}_f \hat{O}_c \nonumber\\
&\times\bigg \{& \vsig_4 \cdot \bigg[\left(\frac{\textbf{p}_5}{2m_5}+\frac{\textbf{p}_4}{2m_4}\right)
-\left(\frac{\textbf{p}_3'}{2m_3'}+\frac{\textbf{p}_3}{2m_3}\right)
+i\vsig_3 \times \left(\frac{\textbf{p}_3}{2m_3}-\frac{\textbf{p}_3'}{2m_3'}\right)
\bigg]\nonumber\\
&+&
\vsig_3 \cdot \bigg[\left(\frac{\textbf{p}_3'}{2m_3'}+\frac{\textbf{p}_3}{2m_3}\right)
-\left(\frac{\textbf{p}_5}{2m_5}+\frac{\textbf{p}_4}{2m_4}\right)
+i \vsig_4\times \left(\frac{\textbf{p}_4}{2m_4}-\frac{\textbf{p}_5}{2m_5}\right)
\bigg]\bigg \}\,,\nonumber \\
O_{\rm w}^{PV}&=&
\delta^3(\textbf{p}_3-\textbf{p}_3'-\textbf{p}_4-\textbf{p}_5)/(2\pi)^3
\hat{O}_f \hat{O}_c (\vsig_3 \cdot \vsig_4 - 1)\,,
\end{eqnarray}
where $\textbf{p}_j$, $m_j$, and $\vsig_j$ as
the momentum, mass,  and spin of the $j$-quark, respectively,
are assigned as in Fig.~\ref{tu}.
In Eq.~(\ref{Hw}), $\hat{O}_f= b_5^\dagger(d)b_4^\dagger(\bar{u})b_3^\dagger(c)b_3(b)$
transforms $b$  to $c$ quark for the $\Omega_b^-\to \Omega_c^{(*)}$ transition,
and creates the $\bar ud$ pair for the final-state meson.
Moreover, $\hat{O}_c=\delta_{c_4,c_5}\delta_{c_3',c_3}$ is the color operator,
where $\delta_{c_m,c_n}$ corresponds to the $W$-boson emission,
which is color-singlet.

In the constituent quark model,
the meson and baryon wave functions are established as
\begin{eqnarray}\label{mock-meson}
| M(\mathbf{P}_M,J,J_z)\rangle&=&\sum_{S_z,L_z}\langle L,L_z;S,S_z |J,J_z \rangle
\int d  \mathbf{p}_4 d  \mathbf{p}_5 \delta^3( \mathbf{p}_4 + \mathbf{p}_5- \mathbf{P}_M)\nonumber\\
&\times&
\Psi_M(\mathbf{p}_4,\mathbf{p}_5)|q_4(\mathbf{p}_4) q_5(\mathbf{p}_5)\rangle\,,\nonumber\\
|{\bf B}( \mathbf{P}_{\bf B},J,J_z)\rangle&=&\sum_{S_z,L_z}\langle L,L_z;S,S_z |J,J_z \rangle\int d \mathbf{p}_1 d \mathbf{p}_2 d \mathbf{p}_3 \delta^3( \mathbf{p}_1 + \mathbf{p}_2+ \mathbf{p}_3- \mathbf{P}_{\bf B})\,\nonumber\\
&\times&\Psi_{\bf B}(\mathbf{p}_1,\mathbf{p}_2,\mathbf{p}_3)
|q_1(\mathbf{p}_1) q_2(\mathbf{p}_2) q_3(\mathbf{p}_3)\rangle\,,
\end{eqnarray}
which correspond to the kets $|M\rangle$ and $|\Omega_b\rangle$ in Eq.~(\ref{dww3}),
respectively. By adding a prime notation in $\mathbf{p}_3$, $|{\bf B}( \mathbf{P}_{\bf B},J,J_z)\rangle$
correspond to $\Omega_c^{(*)0}$.
The meson wave function $\Psi_M(\mathbf{p}_4,\mathbf{p}_5)$ in the momentum space
is given by
\begin{eqnarray}\label{MWF}
\Psi_{{M}}(\textbf{p}_4,\textbf{p}_5)=\zeta_{{M}}\varphi_{{M}}\chi^S_{S_z}\psi_M(\textbf{p}_4,\textbf{p}_5)\,,
\end{eqnarray}
where $\zeta_{M}$, $\varphi_{M}$, and $\chi^S_{S_z}$
stand for the color, flavor, and spin wave functions, respectively.
For the final-state mesons $\pi^-$ and $\rho^-$,
we present $\zeta_{\pi^-,\rho^-}=(R\bar{R}+G\bar{G}+B\bar{B})/\sqrt{3}$
and $\varphi_{\pi^-,\rho^-} = \bar{u}d$, while
$\chi^0_{0}=(\uparrow\downarrow-\downarrow\uparrow)/\sqrt{2}$ and $\chi^1_{1,0,-1}=
[\uparrow\uparrow,(\uparrow\downarrow+\downarrow\uparrow)/\sqrt{2},\downarrow\downarrow]$
are the $\pi^-$ and $\rho^-$ spin wave functions, respectively.
As the spatial wave function, $\psi_M(\textbf{p}_4,\textbf{p}_5)$
is written as
\begin{eqnarray}\label{3wf}
\psi_{\pi(\rho)}(\textbf{p}_4,\textbf{p}_5)=\frac{2^{3/4}}{\pi^{3/4}\beta_{\pi(\rho)}^{3/2}}\mathrm{exp}
\left[-\frac{(\textbf{p}_4-\textbf{p}_5)^2}{4\beta_{\pi(\rho)}^2}\right]\,,
\end{eqnarray}
where the coefficient $\beta_{\pi(\rho)}$ controls the Gaussian distribution.

The baryon wave function $\Psi_{\bf B}(\mathbf{p}_1,\mathbf{p}_2,\mathbf{p}_3)$
can be more complicated, given by
\begin{eqnarray}
\Psi_{\bf B}(\mathbf{p}_1,\mathbf{p}_2,\mathbf{p}_3)=
\zeta_{\bf B} \varphi_{\bf B}\chi^S_{S_z}\Psi_{NLM_L}(\mathbf{p}_1,\mathbf{p}_2,\mathbf{p}_3)\,,
\end{eqnarray}
where $\zeta_{\bf B}$, $\varphi_{\bf B}$, $\chi^S_{S_z}$, and $\Psi_{NLM_L}$
represent the color, flavor, spin, and spatial wave functions, respectively.
For the $\Omega_{b(c)}$ and $\Omega_{c}^{(*)}$ baryons, the corresponding color and flavor wave functions
are $\zeta_{\Omega_{b}}=\zeta_{\Omega_{c}^{(*)}} =(RGB-RBG+GBR-GRB+BRG-BGR)/\sqrt{6}$ and
$\varphi_{\Omega_b,\Omega_c^{(*)}} = (bss,css)$.

The spin and spatial wave functions of a baryon are not independent,
since both are determined by the internal three-quark dynamics.
To illustrate this relation, one may adopt the Jacobi-momentum framework,
in which the three-momenta $\mathbf{p}_{1,2,3}$ of the constituent quarks $q_{1,2,3}$
are reorganized as
\begin{eqnarray}
\mathbf{p}_\rho=\frac{\sqrt{2}}{2}(\textbf{p}_1-\textbf{p}_2)\,,\;
\mathbf{p}_\lambda=
\frac{\sqrt{6}}{2}\frac{m_3(\textbf{p}_1+\textbf{p}_2)-(m_1+m_2)\textbf{p}_3}{m_1+m_2+m_3}\,.
\end{eqnarray}
Here, $\mathbf{p}_\rho$ describes the relative motion between $q_1$ and $q_2$, while
$\mathbf{p}_\lambda$ characterizes the motion of the third quark $q_3$
with respect to the $q_1$-$q_2$ subsystem. Accordingly,
the $q_1q_2$ pair forms a diquark associated with the $\rho$-mode oscillator,
whereas the motion between $q_3$ and the diquark corresponds to the $\lambda$-mode oscillator.
The baryon wave function, originally written as
$\chi^S_{S_z,s_\rho}\Psi_{NLM_L}(\mathbf{p}_1,\mathbf{p}_2,\mathbf{p}_3)$,
can therefore be expressed in the Jacobi basis as
$\chi^S_{S_z,s_\rho}\Psi_{NLM_L}(\mathbf{p}_\rho,\mathbf{p}_\lambda)$,
where the $\rho$ label indicates that the spin structure of the $q_1q_2$ subsystem
is correlated with the corresponding $\rho$-mode spatial excitation.

The $\Omega_b$ and the ground-state $\Omega_c$ baryons are both spin-1/2 fermions.
In these systems, the two strange quarks form a symmetric $ss$-diquark with spin $s_\rho=1$,
leading to the total spin $S=s_\rho-s_c=1/2$, where $s_c=1/2$ denotes the charm quark spin.
Accordingly, their spin wave functions can be written as $\chi^{1/2}_{1/2,1}=
-1/\sqrt{6}(\uparrow\downarrow\uparrow+\downarrow\uparrow\uparrow-2\uparrow\uparrow\downarrow)$
and $\chi^{1/2}_{-1/2,1}=1/\sqrt{6}(\uparrow\downarrow\downarrow+\downarrow\uparrow\downarrow-2\downarrow\downarrow\uparrow)$.
As another example, $\Omega_c(2770)$ also contains a spin-1 $ss$ diquark ($s_\rho=1$)
but carries total spin $S=s_\rho+s_c=3/2$. Its spin wave-function components
are given by $\chi^{3/2}_{3/2,1}=\uparrow\uparrow\uparrow$,
$\chi^{3/2}_{1/2,1}=
(\uparrow\uparrow\downarrow+\uparrow\downarrow\uparrow+\downarrow\uparrow\uparrow)/\sqrt{3}$,
$\chi^{3/2}_{-1/2,1}=1/\sqrt{3}(\uparrow\downarrow\downarrow+\downarrow\uparrow\downarrow+\downarrow\downarrow\uparrow)$, and
$\chi^{3/2}_{-3/2,1}=\downarrow\downarrow\downarrow$.

%
\begin{table}[b]
\begin{center}
\caption{The spin-space wave functions of the $\Omega_c$ baryons up to the $1D$-wave states,
along with the assignment of the quantum numbers.
The configurations are labeled in the form
$|N^{2S+1}L J^P\rangle$.}\label{tab0}\setlength{\tabcolsep}{1.5mm}
{
\scriptsize
\begin{tabular}{cccccccccccccccc}\hline\hline
& $\Omega_c$ baryon states
& $N$
& $n_\rho$
& $n_\lambda$
& $L$
& $S$
& $s_\rho$
& $s_c$
& $l_{\rho}$
& $l_{\lambda}$
& $J^{P}$
& Wave function         \\

\hline
& $|\Omega_c ~1^{4}S\frac{3}{2}^{+}\rangle$
& $1$
& 0
& 0
& 0
& $\frac{3}{2}$
& $1$
& $\frac{1}{2}$
& $0$
& $0$
& $\frac{3}{2}^{+}$
& $\chi^{3/2}_{S_z,1} \psi_{000}(\textbf{p}_\rho)\psi_{000}(\textbf{p}_\lambda) $\\

& $|\Omega_c ~1^{2}S\frac{1}{2}^{+}\rangle$
& $1$
& 0
& 0
& 0
& $\frac{1}{2}$
& $1$
& $\frac{1}{2}$
& $0$
& $0$
& $\frac{1}{2}^{+}$
& $\chi^{1/2}_{S_z,1} \psi_{000}(\textbf{p}_\rho)\psi_{000}(\textbf{p}_\lambda) $\\

& $|\Omega_c ~2^{4}S_{\lambda } \frac{3}{2}^{+}\rangle$
& $2$
& 0
& 1
& 0
& $\frac{3}{2}$
& $1$
& $\frac{1}{2}$
& $0$
& $0$
& $\frac{3}{2}^{+}$
& $\chi^{3/2}_{S_z,1} \psi_{000}(\textbf{p}_\rho)\psi_{100}(\textbf{p}_\lambda) $\\

& $|\Omega_c ~2^{2}S_{\lambda } \frac{1}{2}^{+}\rangle$
& $2$
& 0
& 1
& 0
& $\frac{1}{2}$
& $1$
& $\frac{1}{2}$
& $0$
& $0$
& $\frac{1}{2}^{+}$
& $\chi^{1/2}_{S_z,1} \psi_{000}(\textbf{p}_\rho)\psi_{100}(\textbf{p}_\lambda) $\\

& $|\Omega_c ~1^{4}P_{\lambda }J^{-}\rangle$
& $1$
& 0
& 0
& 1
& $\frac{3}{2}$
& $1$
& $\frac{1}{2}$
& $0$
& $1$
& $\frac{5}{2}^{-},\frac{3}{2}^{-},\frac{1}{2}^{-}$
& $\chi^{3/2}_{S_z,1} \psi_{000}(\textbf{p}_\rho)\psi_{01m_\lambda}(\textbf{p}_\lambda) $\\

& $|\Omega_c ~1^{2}P_{\lambda }J^{-}\rangle$
& $1$
& 0
& 0
& 1
& $\frac{1}{2}$
& $1$
& $\frac{1}{2}$
& $0$
& $1$
& $\frac{3}{2}^{-},\frac{1}{2}^{-}$
& $\chi^{1/2}_{S_z,1} \psi_{000}(\textbf{p}_\rho)\psi_{01m_\lambda}(\textbf{p}_\lambda) $\\

& $|\Omega_c ~1^{4}D_{\lambda }J^{+}\rangle$
& $1$
& 0
& 0
& 2
& $\frac{3}{2}$
& $1$
& $\frac{1}{2}$
& $0$
& $2$
& $\frac{7}{2}^{+},\frac{5}{2}^{+},\frac{3}{2}^{+},\frac{1}{2}^{+}$
& $\chi^{3/2}_{S_z,1} \psi_{000}(\textbf{p}_\rho)\psi_{02m_\lambda}(\textbf{p}_\lambda) $\\

& $|\Omega_c ~1^{2}D_{\lambda }J^{+}\rangle$
& $1$
& 0
& 0
& 2
& $\frac{1}{2}$
& $1$
& $\frac{1}{2}$
& $0$
& $2$
& $\frac{5}{2}^{+},\frac{3}{2}^{+}$
& $\chi^{1/2}_{S_z,1} \psi_{000}(\textbf{p}_\rho)\psi_{02m_\lambda}(\textbf{p}_\lambda) $\\
\hline
\hline
\end{tabular}}
\end{center}
\end{table}
%

In the $\rho$-$\lambda$ oscillation model,
the spatial wave function of a three-quark system is expressed as~\cite{Weng:2024roa}
\begin{eqnarray}\label{rho-lambda-1}
\Psi_{N L L_z}(\textbf{p}_\rho,\textbf{p}_\lambda)
=\sum_{m_\rho, m_\lambda} \langle l_\rho, m_\rho;l_\lambda, m_\lambda|L,L_z \rangle
\psi_{n_\rho l_\rho m_\rho }(\textbf{p}_\rho)
\psi_{n_\lambda l_\lambda m_\lambda }(\textbf{p}_\lambda)\,,
\end{eqnarray}
where the principal quantum number is defined as $N=n_{\rho}+n_{\lambda}+1$,
the total angular momentum satisfies $L=l_\rho+l_\lambda,..., |l_\rho-l_\lambda|$, and
$L_z=m_\rho+m_\lambda$. Here, $(n_i,l_i,m_i$) with $i=\rho$ or $\lambda$
denote the radial, orbital, and magnetic quantum numbers, respectively,
in the $\rho$ or $\lambda$ oscillation system. Assuming the simple harmonic oscillator wave functions,
$\psi_{n_{\rho} l_{\rho} m_{\rho}}(\textbf{p}_{\rho})$ and
$\psi_{n_\lambda l_\lambda m_\lambda }(\textbf{p}_\lambda)$ in Eq.~(\ref{rho-lambda-1})
share the same functional form,
\begin{eqnarray}\label{wf}
\psi_{n l m}(\textbf{p})=(i)^l(-1)^n\left[\frac{2n!}{(n+l+1/2)!}\right]^{1/2}\frac{1}{\alpha^{l+3/2}}
\mathrm{exp}\left(-\frac{\textbf{p}^2}{2\alpha^2}\right)L_n^{l+1/2}(\textbf{p}^2/\alpha^2)\mathcal{Y}_{lm}(\textbf{p})\,,
\end{eqnarray}
where $L_n^{l+1/2}$ is the associated Laguerre polynomial and the solid harmonic is defined by
$\mathcal{Y}_{lm}(\textbf{p})=|\textbf{p}|^{l}Y_{lm}(\mathbf{\hat{p}})$.
The oscillator parameter $\alpha$ corresponds to either $\alpha_\rho$ or $\alpha_\lambda$.
For the $\Omega_b$ and $\Omega_c^{(*)}$ wave functions,
the $\lambda$-mode oscillator parameter is related to $\alpha_\rho$ by
$\alpha_{\lambda}=[(3m_{b(c)})/(2m_s+m_{b(c)})]^{1/4}\alpha_{\rho}$,
as obtained in Refs.~\cite{Wang:2017kfr,Yao:2018jmc}.

With these ingredients, the spin-space wave functions of the $\Omega_c$ baryons
can be constructed, as summarized in Table~\ref{tab0}.
In particular, the notation $|N^{2S+1}LJ^P\rangle$ specifies
the quark configuration of the $\Omega_c$ baryon in the $\rho$-$\lambda$ scheme,
where 
the orbital angular momentum $L=0,1,2$ corresponds to the $S$-, $P$-, and $D$-wave states,
respectively. The total spin is given by $S=s_\rho\pm s_c$, and the total angular momentum
is obtained from $J=L+S$. The subscript $\lambda$ in $S_\lambda$ or $P_\lambda$ ($D_\lambda$)
indicates that the excitation arises from the $\lambda$ mode, corresponding to
$N=n_\lambda+1$ (radial excitation) or $L=l_\lambda$ (orbital excitation) in the $\lambda$ oscillator.
These constructions lead to the spin-space wave functions (see the last column of Table~\ref{tab0}).

In the constituent quark model, the transition amplitude $\langle \Omega_c^{(*)0} M|O_1|\Omega_b^-\rangle$
with the insertion of $O_{\rm w}^{PC,PV}$ in Eq.~(\ref{Hw}) and the wave functions in Eq.~(\ref{mock-meson})
is evaluated as
\begin{eqnarray}\label{ampA}
&&
\mathcal{M}_{J_f,J_f^z;J_{{M}},J_{{M}}^z}^{J_i,J_i^z}(\Omega_b^- \to \Omega_c^{(*)}{M})
= \mathcal{M}_{J_f,J_f^z;J_{{M}},J_{{M}}^z}^{J_i,J_i^z,PC}
+\mathcal{M}_{J_f,J_f^z;J_{{M}},J_{{M}}^z}^{J_i,J_i^z,PV}\,,\nonumber\\
&&
\mathcal{M}_{J_f,J_f^z;J_{{M}},J_{{M}}^z}^{J_i,J_i^z,PC(PV)}=
\langle \Omega_c^{(*)}(\textbf{P}_f,J_f,J_f^z) {M}(\textbf{q},J_{{M}},J_{{M}}^z)|O_{\rm w}^{PC(PV)}|\Omega_b(\textbf{P}_i,J_i,J_i^z)\rangle\,,
\end{eqnarray}
where $J_{i,f,M}^{(z)}$ denote the total (third) angular momenta of the initial-state baryon, final baryon,
and final meson, respectively, and $\textbf{q}=\textbf{P}_i-\textbf{P}_f$ is the momentum transfer.

As an explicit example in our calculation,
we consider the decay $\Omega_b^- \rightarrow \Omega_c^0 \pi^-$.
The corresponding angular-momentum assignments are
$(J_f,J_{{M}},J_i)=(1/2,0,1/2)$ and $(J_f^z,J_{{M}}^z,J_i^z)=(-1/2,0,-1/2)$.
Using Eq.~(\ref{ampA}), the parity-conserving (PC) and parity-violating (PV) amplitudes
are obtained as
\begin{eqnarray}\label{equa1}
\mathcal{M}_{1/2,-1/2,0,0}^{1/2,-1/2,PC} &=&
-\frac{\sqrt{2}}{\sqrt{3}\pi ^{9/4}} a_1 G_F V_{cb} V_{ud}^{\ast} \frac{ T_\alpha \beta_{\pi}^{9/2}}{(\alpha_{\lambda_f}^2+\alpha_{\lambda_i}^2) (2 m_2+m_3')} q  F(q^2) \nonumber\\
&&
\times\bigg[\alpha_{\lambda_i}^2 \left(1-\frac{2m_2}{m_3}+\frac{m_2}{m_4}+\frac{m_2}{m_5}+\frac{m_3'}{2 m_4}+\frac{m_3'}{2 m_5}\right) \nonumber\\
&&+\alpha_{\lambda_f}^2 \left(1+\frac{{m_2}}{m_4}+\frac{m_2}{m_5}+\frac{2 m_2}{m_3'}+\frac{m_3'}{2 m_4}+\frac{m_3'}{2 m_5}\right) \bigg]\,,\nonumber\\
\mathcal{M}_{1/2,-1/2,0,0}^{1/2,-1/2,PV} &=&
-\frac{2\sqrt{6}}{ \pi ^{9/4}}a_1 G_F V_{cb} V_{ud}^{\ast}  \beta_{\pi}^{9/2} T_\alpha F(q^2)\,,
\end{eqnarray}
where the oscillator parameters are defined by
$\alpha_{\lambda_i}= [3m_b/(2m_s+m_b)]^{1/4}\alpha_{\rho_i}$ and
$\alpha_{\lambda_f}=[3m_c/(2m_s+m_c)]^{1/4}\alpha_{\rho_f}$.
The quantities $T_\alpha$ and $F(q^2)$ are given by
\begin{eqnarray}
&&
T_\alpha=
\left[\frac{\alpha_{\lambda_f}\alpha_{\lambda_i}\alpha_{\rho_f}\alpha_{\rho_i}}
{\beta_{\pi}^2 \left(\alpha_{\lambda_f}^2+\alpha_{\lambda_i}^2\right)
\left(\alpha_{\rho_f}^2+\alpha_{\rho_i}^2\right)}\right]^{3/2},
F(q^2)=
\exp \left[\frac{-3 m_2^2 q^2}{\left(\alpha_{\lambda_f}^2+\alpha_{\lambda_i}^2\right)(2 m_2+m_3')^2}\right]\,.
\end{eqnarray}
Using the equation
\begin{eqnarray}\label{dww}
\Gamma &=& 8\pi^2\frac{|\textbf{q}|E_{{M}}E_{f}}{M_{i}}\frac{1}{2J_{i}+1}
\sum_{J_f^z,J_i^z}|\mathcal{M}_{J_f,J_f^z;J_{{M}},J_{{M}}^z}^{J_i,J_i^z}(\Omega_b \to {M} \Omega_c^{(*)})|^2\,,
\end{eqnarray}
the helicity amplitudes are converted into decay widths for the numerical analysis.
Here, $\mathbf{q}$ denotes the three-momentum of the final-state particles in the $\Omega_b$ rest frame,
and the spin projections $J_i^z$ and $J_f^z$ are summed over in the squared amplitudes,
with the factor $1/(2J_i+1)$ accounting for the average over the initial-state polarization.
As a result, the PC and PV amplitudes for the decay
$\Omega_b^- \rightarrow \Omega_c^0\pi^-$,
as constructed in Eq.~(\ref{equa1}), are directly substituted into Eq.~(\ref{dww})
to obtain the corresponding decay width. The same procedure is applied
to compute all other decay widths considered in this work.

\section{Numerical Analysis}
In the numerical analysis, we adopt the CKM matrix elements
$(V_{cb},V_{ud}) = (0.042,0.978)$ from the PDG~\cite{pdg} and the quark masses
$(m_s,m_c,m_b)=(0.55,1.45,4.80)$~GeV from Refs.~\cite{Ni:2023lvx,Wang:2025pcs},
together with the parameter $a_1=1.1$ taken from Ref.~\cite{Gutsche:2018utw,Wang:2024mjw}.
The harmonic oscillator parameters $\alpha_{\rho_i(\rho_f)}$ and $\beta_{\pi(\rho)}$
are taken from Ref.~\cite{Wang:2017kfr} and  Refs.~\cite{Zhong:2008kd,Wang:2025pcs}, respectively,
with values
\begin{eqnarray}\label{para}
&&
\alpha_{\rho_i}=\alpha_{\rho_f}=440~\text{MeV}\,,\;\nonumber\\
&&
\beta_{\pi}=\beta_{\rho}=400~\text{MeV}\,.
\end{eqnarray}
Using the parameters in Eq.~(\ref{para}) as theoretical inputs, together with
$O_{\rm w}^{PC,PV}$ in Eq.~(\ref{Hw}),
the wave functions in Eqs.~(\ref{mock-meson}, \ref{rho-lambda-1}, \ref{wf}), and
the amplitudes in Eq.~(\ref{ampA}), we evaluate the branching fractions
${\cal B}(\Omega_b^- \to \Omega_c^0 \pi^-,\Omega_c^0 \rho^-)$,
where $\Omega_c^0$ is the lowest-lying $1S$-wave $\Omega_c$ state.
The numerical results are presented in Table~\ref{tab1},
together with a comparison with other model predictions.
We further predict the branching fractions for
$\Omega_b \to \Omega_c^{*} M$, including final-state $\Omega_c$ baryons in the
$1S$-, $1P$-, $2S$-, and $1D$-wave configurations.

In the $L$--$S$ coupling scheme,
the $\Omega_c$ wave states are labeled by $|LJ^P\rangle$, as listed in Table~\ref{tab0}.
In this framework, the total orbital angular momentum
$L$ is obtained by coupling the $\rho$- and $\lambda$-mode orbital momenta,
$L={\ell}_\rho+{\ell}_\lambda$, while the total spin is given by $S=s_\rho+s_c$.
The total angular momentum of the charmed baryon $\Omega_c^{(*)}$  is then obtained as $J=L+S$.
It is worth noting that, in the heavy-quark limit,
the charm-quark spin $s_c$ decouples from the light degrees of freedom.
This motivates the use of a more suitable framework,
in which the $\Omega_c$ wave states are labeled by $|J^P,j_\ell\rangle$ in the $j-j$ coupling scheme~\cite{Wang:2017hej,Wang:2018fjm,Xiao:2020oif}.
In this notation, $j_\ell=L+s_\rho$ denotes the total angular momentum of the light-quark system.
The total angular momentum of the baryon is then
obtained by coupling $j_\ell$ with the charm-quark spin, $J=j_\ell+s_c$,
for the $\Omega_c^{(*)}$ states considered in this work.

The two coupling schemes are related by a unitary transformation~\cite{Roberts:2007ni},
\begin{eqnarray}\label{Rela}
|J^P,j_\ell\rangle=\sum_{S}(-1)^{L+s_\rho+s_c+J}\sqrt{(2S+1)(2j_\ell+1)}
\begin{Bmatrix}
s_\rho & s_c & S\\L & J & j_\ell
\end{Bmatrix}
|LJ^P\rangle\,,
\end{eqnarray}
where the bracket denotes the Wigner $6j$ symbol describing the recoupling of angular momenta.
In the $j$–$j$ coupling scheme, the charmed baryons $\Omega_c^{(*)}$
considered in this work are labeled by $|J^P,j_\ell\rangle$, as summarized in Table~\ref{tab2}.
Moreover, the subscript $\lambda$ in $|J^P,j_\ell\rangle_\lambda$
for the $1P$-, $2S$-, and $1D$-wave states
indicates that the excitation occurs in the $\lambda$ mode,
while the $\rho$ mode remains in its ground state.

%
\begin{table}[b]
\begin{center}
\caption{\label{s-wave} Branching fractions of $\Omega_b^-  \rightarrow \Omega_c^0 \pi^-$ and
$\Omega_b^-  \rightarrow \Omega_c^0 \rho^-$, in comparison with previous works.}~\label{tab1}
{
\scriptsize
\begin{tabular}{lccccccccccccccccccccccccccccccccccccccccccccc}\hline\hline
&$10^3 \mathcal{B}(\Omega_b^-\rightarrow \Omega_c^0 \pi^-)$
&$10^3 \mathcal{B}(\Omega_b^-\rightarrow \Omega_c^0 \rho^-)$
&\\
\hline
CQM (this work)                                 &1.2           &6.3                        \\
LFQM~\cite{Chua:2019yqh}                    &$1.1^{+0.85}_{-0.55}$       &$3.07^{+2.41}_{-1.53}$                      \\
LFQM~\cite{Zhao:2018zcb}                    &4.0           &10.8                       \\
LFQM~\cite{Li:2021kfb}                      &2.82          &7.92                        \\
QCDF~\cite{Neishabouri:2026mjn}             &1.809         &5.047                       \\
RQM~\cite{Ivanov:1997ra,Ivanov:1997hi}      &5.81          &$\cdots$                    \\
RQM~\cite{Patel:2025gbw}                    &2.54          &7.43                       \\
CCQM~\cite{Gutsche:2018utw}                 &1.88          &5.4                        \\
NRQM~\cite{Cheng:1996cs}                    &4.92          &17.9                      \\
\hline\hline
\end{tabular}}
\end{center}
\end{table}
%

\section{Discussions and Conclusion}
Experimentally, the decay $\Omega_b^-\to \Omega_c^0\pi^-$ has been observed
through its subsequent decay $\Omega_c^0\to p K^- K^-\pi^+$~\cite{LHCb:2016coe}.
However, the corresponding branching fraction has not yet been reported. Consequently,
experimental data are currently insufficient to test the validity of theoretical models
for processes involving the ground-state $\Omega_c^0$. On the other hand,
as shown in Table~\ref{tab1}, our predicted branching fractions
${\cal B}(\Omega_b^- \to \Omega_c^0 \pi^-,\Omega_c^0 \rho^-)=(1.2,6.3) \times 10^{-3}$
are consistent, at least at the order-of-magnitude level, with the existing calculations
in the light-front quark model~(LFQM)~\cite{Chua:2019yqh,Zhao:2018zcb,Li:2021kfb},
the relativistic quark model (RQM)~\cite{Ivanov:1997ra,Ivanov:1997hi,Patel:2025gbw},
the QCD factorization approach~\cite{Patel:2025gbw},
and the covariant confined quark model (CCQM)~\cite{Gutsche:2018utw}.
This agreement suggests that the constituent quark model employed in this work
provides a reasonable framework comparable to other theoretical approaches.
Based on this consistency, we further present our predictions
for $\Omega_b^-\to\Omega_c^{*0}\pi^-,\Omega_c^{*0}\rho^-$.

In the $1S$-wave classification, there exists an excited state, $\Omega_c(2770)^0$, with spin $3/2$.
Its existence has been experimentally established, and it is found to decay dominantly through the radiative channel.
Using the branching fractions
${\cal B}(\Omega_b^-\to \Omega_c(2770)^0\pi^-)$ and ${\cal B}(\Omega_b^-\to \Omega_c(2770)^0\rho^-)$
predicted in Table~\ref{tab2}, together with
${\cal B}(\Omega_c(2770)^0\to \Omega_c^0\gamma)\simeq 100\%$ from the PDG~\cite{pdg},
we estimate the cascade radiative branching fractions
\begin{eqnarray}\label{rad}
{\cal B}(\Omega_b^-\to \Omega_c^0\pi^-\gamma,\Omega_c^0\rho^-\gamma)
\simeq (5.2,1.4)\times 10^{-3}\,.
\end{eqnarray}
In addition, radiative processes $\Omega_b^- \to \Omega_c^0\pi^-\gamma$ and
$\Omega_b^- \to \Omega_c^0\rho^-\gamma$ can also arise from photon emission via
inner bremsstrahlung (IB) from the charged external legs in the non-radiative decays
$\Omega_b^- \to \Omega_c^0\pi^-$ and $\Omega_b^- \to \Omega_c^0\rho^-$.
Since the IB contribution is infrared divergent,
we implicitly assume a photon-energy threshold
$E_\gamma>E_0$ in the $\Omega_b$ rest frame.
A rough estimate gives ${\cal B}_{\rm IB}\sim \delta_{\rm IB}\,{\cal B}$ with
$\delta_{\rm IB}={\cal O}\!\left[(\alpha_{\rm em}/\pi)\ln(E_{\rm hard}/E_0)\right]\sim 10^{-2}$~\cite{Geng:2004jk}
for typical experimental cuts, leading to
${\cal B}_{\rm IB}(\Omega_b^- \to \Omega_c^0\pi^-\gamma,\Omega_c^0\rho^-\gamma)\sim 10^{-5}$.
By comparison, the cascade contributions in Eq.~(\ref{rad}) are significantly larger than the IB background,
indicating that these radiative $\Omega_b$ decay modes may provide promising channels for future experimental measurements.

%
\begin{table}[b]
\begin{center}
\caption{Branching fractions of $\Omega_b\to\Omega_c^{(*)} {M}$ decay processes.
The final-state $\Omega_c^{(*)}$ are classified, from top to button,
as the $1S$-, $1P$-, $2S$-, and $1D$-wave states. The corresponding masses are
denoted by $M_f$, while $|J^P,j_\ell\rangle$ specifies the wave state of
the $\Omega_c^{(*)}$ baryon. We define
${\cal B}_M={\cal B}(\Omega_b\to\Omega_c^{(*)} M)$
to present our predicted branching fractions.}\label{tab2}
{
\scriptsize
\begin{tabular}{c|lcccccccccccccccccccccccccccccccccccccccccc}\hline\hline
wave
&$\Omega_c$ baryons
&$|J^P,j_\ell\rangle$
&$N$
&$n_{\lambda}$
&$n_{\rho}$
&$l_{\lambda}$
&$l_{\rho}$
&$L$ &$s_{\rho}$
&$j_\ell$
&$J^P$
&$M_f$ (MeV)
&$10^3 \mathcal{B}_\pi$
&$10^3 \mathcal{B}_\rho$ \\ \hline
1S&$\Omega_c^0$
&${|{J^P=\frac{1}{2}^+,1}\rangle}$
&1 &0  &0  &0  &0  &0  &1  &1  &$\frac{1}{2}^{+}$       &2695         &1.2 &6.3\\
&$\Omega_c(2770)$
&${|{J^P=\frac{3}{2}^+,1}\rangle}$
&1 &0  &0  &0  &0  &0  &1  &1  &$\frac{3}{2}^{+}$       &2770         &5.2  &1.4\\
\hline
1P&$\Omega_c^{\prime}(3000)$ 
&${|{J^P=\frac{1}{2}^-,0}\rangle}_\lambda$
&1 &0  &0  &1  &0  &1  &1  &0  &$\frac{1}{2}^{-}$       &3000          &2.2   &2.0\\
&$\Omega_c(3050)$
&${|{J^P=\frac{3}{2}^-,1}\rangle}_\lambda$
&1 &0  &0  &1  &0  &1  &1  &1  &$\frac{3}{2}^{-}$       &3055          &1.9   &2.3\\
&$\Omega_c(3000)$
&${|{J^P=\frac{1}{2}^-,1}\rangle}_\lambda$
&1 &0  &0  &1  &0  &1  &1  &1  &$\frac{1}{2}^{-}$       &3000          &4.4   &4.0\\
&$\Omega_c(3065)$
&${|{J^P=\frac{3}{2}^-,2}\rangle}_\lambda$
&1 &0  &0  &1  &0  &1  &1  &2  &$\frac{3}{2}^{-}$       &3066           &1.3   &9.8\\
&$\Omega_c(3090)$
&${|{J^P=\frac{5}{2}^-,2}\rangle}_\lambda$	
&1 &0  &0  &1  &0  &1  &1  &2  &$\frac{5}{2}^{-}$       &3090           &8.1   &1.5\\
\hline
2S&$***$ 
&${|{J^P=\frac{1}{2}^+,1}\rangle}_{\lambda}$
&2 &1  &0  &0  &0  &0  &1  &1  &$\frac{1}{2}^{+}$       &3088           &1.5   &7.3\\
&$\Omega_c(3185)$
&${|{J^P=\frac{3}{2}^+,1}\rangle}_{\lambda}$	
&2 &1  &0  &0  &0  &0  &1  &1  &$\frac{3}{2}^{+}$       &3177           &6.0   &1.4\\
\hline
1D&$***$ 
&${|{J^P=\frac{3}{2}^+,1}\rangle}_{\lambda}$
&1 &0  &0  &2  &0  &2  &1  &1  &$\frac{3}{2}^{+}$       &3296~\cite{Ebert:2011kk}       &0.6   &2.0\\
&$***$ 
&${|{J^P=\frac{5}{2}^+,2}\rangle}_{\lambda}$
&1 &0  &0  &2  &0  &2  &1  &2  &$\frac{5}{2}^{+}$       &3305~\cite{Ebert:2011kk}       &1.5    &1.6\\
&$***$ 
&${|{J^P=\frac{1}{2}^+,1}\rangle}_{\lambda}$
&1 &0  &0  &2  &0  &2  &1  &1  &$\frac{1}{2}^{+}$       &3264~\cite{Ebert:2011kk}       &7.5    &0.3\\
&$***$ 
&${|{J^P=\frac{3}{2}^+,2}\rangle}_{\lambda}$
&1 &0  &0  &2  &0  &2  &1  &2  &$\frac{3}{2}^{+}$       &3280~\cite{Ebert:2011kk}       &3.0     &2.6 \\
&$\Omega_c(3327?)$
&${|{J^P=\frac{5}{2}^+,3}\rangle}_{\lambda}$
&1 &0  &0  &2  &0  &2  &1  &3  &$\frac{5}{2}^{+}$       &3296~\cite{Ebert:2011kk}       &2.0     &5.7\\
&$\Omega_c(3327?)$
&${|{J^P=\frac{7}{2}^+,3}\rangle}_{\lambda}$
&1 &0  &0  &2  &0  &2  &1  &3  &$\frac{7}{2}^{+}$       &3306~\cite{Ebert:2011kk}       &4.6      &0.8\\
\hline\hline
\end{tabular}}
\end{center}
\end{table}
%
Four of the five observed $\Omega_c^{**}$ states can be classified as the $1P$-wave excitations.
In particular, $\Omega_c(3000)$, $\Omega_c(3050)$, $\Omega_c(3065)$, and $\Omega_c(3090)$
are observed as resonant peaks in the $\Xi_c^+ K^-$ invariant mass spectrum from
the cascade process $\Omega_b^- \to \Omega_c^{**}\pi^-$ followed by
$\Omega_c^{**}\to \Xi_c^+K^-$~\cite{LHCb:2021ptx}.
By contrast, $\Omega_c(3120)$ has not yet been experimentally confirmed.
Motivated by its non-observation, $\Omega_c(3120)$ may be interpreted as a state
dominated by a $\rho$-mode excitation~\cite{Zhong:2025oti,Weng:2024roa}.
Since the initial $\Omega_b$ state is characterized by $n_\rho=\ell_\rho=0$ and therefore
cannot efficiently undergo a weak transition to $\rho$-mode excitations,
it is reasonable that the $\Omega_b$ decay into $\Omega_c(3120)$ is forbidden or highly suppressed.

Despite the experimental observation of the processes
$\Omega_b^- \to \Omega_c^{**0}\pi^-$, their branching fractions have not yet been reported.
We therefore present our predictions, together with those for
${\cal B}(\Omega_b^-\to \Omega_c^{**0}\rho^-)$, as listed in Table~\ref{tab2}.
In particular, we highlight several features of our predictions for the $1P$-wave $\Omega_c$ states.
First, the predicted branching fractions
${\cal B}(\Omega_b^-\to\Omega_c(3065)^0\rho^-)$ and
${\cal B}(\Omega_b^-\to\Omega_c(3090)^0\pi^-)$ are of order $10^{-2}$,
which are about five and eight times larger than their corresponding
$\pi$- and $\rho$-mode counterpart channels, respectively. Such significant differences
can serve as useful tests of the constituent quark model (CQM) employed in this work.
Second, the $1P$-wave $\Omega_c$ state $|J^P=1/2^-,0\rangle_\lambda$ is denoted by
$\Omega'_c(3000)$ in Table~\ref{tab2}, as it is predicted to have a mass close to that of
$\Omega_c(3000)$. We find that ${\cal B}(\Omega_b^-\to\Omega'_c(3000)^0\pi^-)$ is comparable
to those of other $1P$-wave states. However,
since the decay width of $\Omega'_c(3000)^0$ is expected to be very broad,  $\Gamma=80-120$ MeV~\cite{Zhong:2025oti,Wang:2017hej}, in the resonant process
$\Omega_b^-\to\Omega'_c(3000)^0\pi^-$ followed by $\Omega'_c(3000)^0\to\Xi_c^+K^-$,
the corresponding resonance may appear broad and flat,
making it difficult to observe in the $\Xi_c^+K^-$ invariant mass spectrum.

Apart from the five $\Omega_c^{**}$ baryons,
the LHCb collaboration reported the two excited $\Omega_c^*$ states
$\Omega_c(3185)$ and $\Omega_c(3327)$ in the proton-proton collision~\cite{LHCb:2023sxp},
not confirmed in the $e^+ e^-$ collision by Belle yet. This leaves the room for
the examinations in $\Omega_b$ weak decays. Thus, we obtain the branching fractions of $\Omega_b^-\to \Omega_c(3185)^0\pi^-$ and $\Omega_b^-\to \Omega_c(3185)^0\rho^-$ to be $(6.0, 1.4)\times 10^{-3}$, respectively. The $\Omega_c(3185)$ baryon is commonly interpreted as a $2S$-wave state with quantum numbers $J^P=3/2^+$~\cite{Zhong:2025oti,Yu:2023bxn,Karliner:2023okv,Pan:2023hwt,Kucukyilmaz:2025rsd}, although alternative interpretations have also been proposed~\cite{Xin:2023gkf,Yan:2023ttx,Yan:2023tvl,Ozdem:2023okg,Feng:2023ixl}.

On the other hand, the theoretical investigations of
Refs.~\cite{Zhong:2025oti,Wang:2023wii,Luo:2023sra,Pan:2023hwt,
Yu:2023bxn,Garcia-Tecocoatzi:2024aqz,Li:2024zze,Weng:2024roa}
interpret the $\Omega_c(3327)$ baryon observed in Ref.~\cite{LHCb:2023sxp}
as a candidate $1D$-wave excitation, leaving
its spin-parity quantum numbers $J^P$ undetermined.
Weak decays of the $\Omega_b$ baryon, such as $\Omega_b^-\to\Omega_c(3327)^0\pi^-$ and
$\Omega_b^-\to\Omega_c(3327)^0\rho^-$, may provide a useful probe
for identifying the quantum numbers of $\Omega_c(3327)^0$. In our framework,
six $1D$-wave $\Omega_c$ states are listed in Table~\ref{tab2}.
Among them, the states $|J^P=5/2^+,3\rangle_{\lambda}$ and
$|J^P=7/2^+,3\rangle_{\lambda}$, whose predicted masses are close to $3327~\text{MeV}$,
are plausible candidates for the newly observed resonance.
Assuming $\Omega_c(3327)$ to be a $1D$-wave excitation with $J^P=5/2^+$ or $7/2^+$~\cite{Zhong:2025oti,Luo:2023sra},
we predict
\begin{eqnarray}
{\cal B}(\Omega_b^-\to \Omega_c(3327)^0\pi^-,\Omega_c(3327)^0\rho^-)
=(2.0,5.7)\times 10^{-3}~\text{or}~(4.6,0.8)\times 10^{-3}\,.
\end{eqnarray}
The pronounced differences between these two scenarios provide a clear discriminant
that can be tested in future measurements at LHCb.
As a final remark,
one and four $\Omega_c^*$ baryons in the $2S$ classification and
the $1D$ classification listed in Table~\ref{tab2}, respectively, remain unexplored
in both scattering processes and weak decays. The branching fractions predicted in this work
therefore provide useful guidance for future experimental searches.

In summary, we have carried out a systematic study of the $\Omega_c$ spectroscopy
up to the $1D$ excitations by analyzing the non-leptonic decays of the $\Omega_b^-$ baryon
within the constituent quark model framework. The predicted branching fractions for
$\Omega_b^- \to \Omega_c^0 \pi^-$ and $\Omega_b^- \to \Omega_c^0 \rho^-$,
at the level of $10^{-3}$, have been found to be consistent with existing theoretical estimates.
Four of the five $\Omega_c^{**}$ states, $\Omega_c(3000)$, $\Omega_c(3050)$,
$\Omega_c(3065)$, and $\Omega_c(3090)$, have been experimentally observed
in $\Omega_b$ decays; however, their production rates have not yet been measured.
We therefore interpret these states as members of the $1P$-wave multiplet
and provide predictions for the branching fractions of
$\Omega_b^- \to \Omega_c^{**}\pi^-,\Omega_c^{**}\rho^-$ at the order of $10^{-3}$.
We have further investigated the newly observed $\Omega_c(3327)$ state
by considering possible $1D$-wave assignments with $J^P=5/2^+$ and $7/2^+$.
The predicted branching fractions for $\Omega_b^- \to \Omega_c(3327)^0\pi^-$ and $\Omega_b^- \to \Omega_c(3327)^0\rho^-$ exhibit distinct patterns
for the two scenarios, providing a sensitive probe of the spin-parity assignment
and internal structure of this state.
These predictions can be tested in future measurements at LHCb.

\section*{Acknowledgements}
KLW was supported by the National Natural Science Foundation of
China (Grants No.~12205026), and Support Project for Young Teachers' Research and Innovation Abilities under Grant No.2025Q037.
YKH was supported by the National Natural Science Foundation of
China (Grants~No.~12575101 and No.~12175128).
JW was supported by Shanxi Provincial Graduate Education Innovation Program Project (Grant~No.~2025XS111).

\end{document}